\documentclass[prb, showpacs, twocolumn] {revtex4}
\usepackage{graphicx}
\begin{document}

\title{A new paradigm for modulated phases in multi-stage 
structural transformations} 

\author{T. Cast\'an$^a$, A. Planes$^a$, and A. Saxena$^{a,b}$ 
\\$^a$Department d'Estructura i Constituents de la Mat\`eria, 
Universitat de Barcelona, Diagonal 647, 08028 Barcelona, Spain   
\\$^b$Theoretical Division, Los Alamos National Laboratory, 
Los Alamos, New Mexico 87545 }     

\date{July 17, 2002} 

\begin{abstract}{For multi-stage, displacive structural transitions 
we present 
a general framework that accounts for various intermediate modulated phases, 
elastic constant, phonon and related thermodynamic anomalies.  Based on the 
presence or absence of modulated phases we classify these transformations in 
four categories and apply this approach to four different representative 
materials: Ni-Mn-Ga, NiTi(Fe),Ni-Al, Cu-Zn-Al and $\alpha$-U. We suggest that 
the anomalous increase in 
elastic constant(s) and phonon frequency observed when approaching the
martensitic transition from above  
is a signature of the conmensurate modulated phase.} 
\end{abstract}
 
\maketitle
 
\section{Introduction}    
 
A quite varied structural and thermodynamic behavior is observed in a
wide class of materials of technological interest such as martensites 
and shape memory alloys \cite{sma}.  A displacive structural transition 
to the low symmetry (`martensitic') phase is often preceded by one or 
more modulated ``phases" \cite{cowley}.  However, under the vast 
experimental data, various anomalies and disparate mechanisms, there 
must be some unifying principles that are common to most of these 
materials. The present study is an attempt to address this question 
by providing a common framework for the different multi-stage 
transformation mechanisms in displacive martensitic transitions (MT). 
This general framework emerges naturally after appropriately 
assimilating the existing experimental data.  The discussion below is
restricted to displacive (non-reconstructive) transitions. 
 
The modulated phase is, in general, an incommensurate phase which, 
eventually, may lock into a commensurate modulation due to the 
freezing of a specific phonon, usually with an associated wavevector 
inside the Brillouin zone. To be more precise, the commensurate 
modulation may or may not be observed but never without 
a preceding incommensurate modulation. A general scenario for a cubic 
symmetry parent (`austenite') phase is depicted in Fig. 1.  
Note that, in general, there may be three transition temperatures: 
(i) $T_I$ is the temperature at which the incommensurate modulation 
(IC) first appears; this is (presumably) a second order transition. 
We note that, in real space, this phase is sometimes referred to as 
`tweed' in the literature. It has been observed in other materials 
\cite{bkwt} such as quartz, high T$_c$ superconducting perovskites, 
ferroelectrics, etc.  (ii) $T_{II}$ is the temperature at which the 
commensurate modulation (CM) appears from a (previous) IC phase; 
this is usually a first order transition. (iii) $T_M$ is the first 
order MT temperature.  Depending on the material, either the 
commensurate phase or both the commensurate and incommensurate phases 
may not appear.  In the latter case there is no modulation or {\it 
precursor} \cite{precursors} phenomena (either in pure or intermetallic 
crystals) and a direct transition to the martensitic phase takes place. 

Concerning the IC phase (`tweed' in real space) we note that in non-stoichiometric 
alloys there could be {\it premonitory} effects such as thermal 
expansion anomalies \cite{jak} that are caused by composition 
fluctuations \cite{kartha95} and stabilized by long-range elastic 
forces \cite{shenoy}.

Based on several experimental observations in a variety of materials 
we first specify the (likely) requirements for a multi-stage transition. 

{\bf (a)} Existence of low restoring forces along specific directions.  In 
most (cubic) martensitic materials (of interest here) this is 
accomplished by a temperature softening of the long-wavelength limit 
of the [110][1$\overline{1}$0] transverse acoustic (TA$_2$) phonon 
branch (i.e., $\Sigma_4$ branch).  This implies that the shear modulus 
$C'=(1/2)(C_{11}-C_{12})$ is smaller than the other elastic constants.  
This effect is related to {\it strain}. 

{\bf (b)} Phonon softening (dip) observed in [110][1$\overline{1}$0]  
TA$_2$ branch inside the Brillouin Zone usually at a particular value 
of the wavevector $q\ne0$.  This has an effect on {\it shuffle} (i.e., 
intracell distortion) modes. 

Hereafter, we shall take these two requirements as necessary (but not
sufficient) conditions for applying the point of view adopted in the 
present study. Besides, they are observed in most martensitic materials 
\cite{planes01} and accepted to be premonitory indications of 
the low temperature martensitic phase. Nevertheless, 
at the MT, neither the elastic constant nor the phonon soften completely. 
It is now accepted that an incomplete softening is enough to drive the 
transition \cite{jak92}. This is because of an interplay between the 
{\it strain} and the {\it shuffle} modes.  For onset of the IC, we assume 
the {\it shuffle} mode(s) to be the primary order parameter (OP).  Symmetry 
allowed {\it strain} component(s) may then couple to the {\it shuffle}
as a secondary order parameter, particularly in the CM phase.  
%However, for the MT, {\it strain} serves as a primary order parameter.  
Depending on the material there may be competition or cooperation 
between the two effects associated with requirements (a) and (b). In 
this sense, it is not clear how the interplay between the two partial 
soft modes operates. In addition, it may be modified by coupling to 
other physical variables such as magnetism (or inhomogeneities).

This paper is organized as follows.  In the next section we provide 
a new classification of martensitic materials in terms of presence 
or absence of modulated phases and associated anomalies in phonon dispersion 
and elastic constants.  Section III contains an effective, coupled 
strain-shuffle, one-dimensional model that explains the observed 
phonon driven anomalies and modulated phases.  In Sec. IV we summarize 
our main findings and propose new experiments that may validate our 
predictions.

\begin{center}
\begin{figure}[th]
\includegraphics[width=7.50cm,clip]{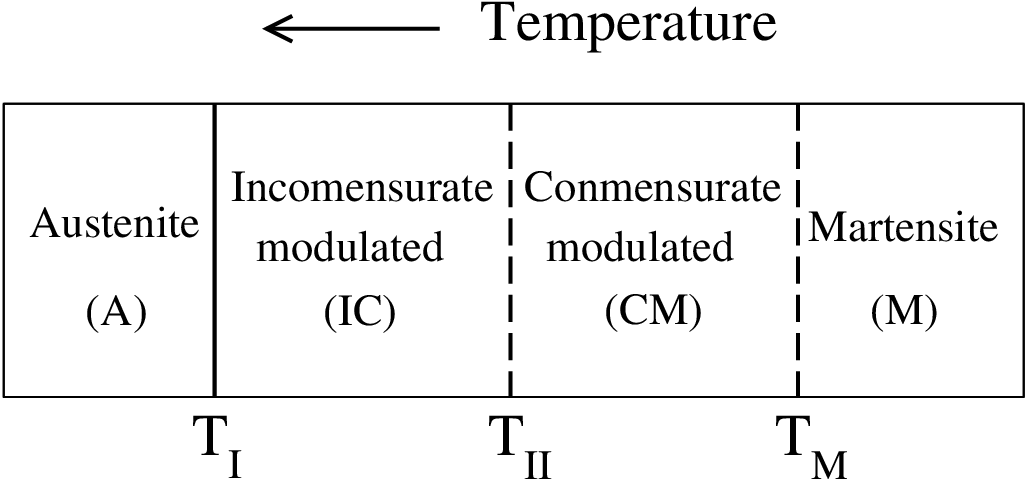}
\caption{Paradigm of the multi-stage displacive structural 
transitions with modulated precursor phases.}
\end{figure}
\end{center}

\section{Representative materials and Observed Anomalies} 
According to the general scenario described above, we can broadly classify 
various displacive martensitic materials in four categories. 

{\bf (I)} Systems that exhibit both, the IC and CM, intermediate transitions. 
The Ni-Mn-Ga (Refs. 10-13) and 
$\alpha$-Uranium (Refs. 14,15) materials belong to this category. 
Both show anomalies in specific heat \cite{lashley,planes97,manosa97}
magnetic susceptibilities \cite{obrado98}, resistivity \cite{lashley}  
and an increase in $C'$ with decreasing temperature between T$_{II}$ 
and T$_M$ \cite{manosa97,stenger98,lander}.

Ni$_2$MnGa is a ferromagnetic  Heusler alloy with cubic symmetry and 
the magnetism (mainly localized on Mn atoms) arises from d-electrons.   
At low temperatures it transforms (for compositions close to stoichiometry) to a tetragonal martensitic phase 
\cite{martynov}.  The phonon anomaly is observed at a wave vector 
$q=\frac{1}{3}[\xi,\xi,0]$, where $0<\xi<1$.

$\alpha$-U is an {\it orthorhombic} material and displays two IC 
phases--incommensurate modulation in one direction or in two directions.  
The origin of magnetism is (likely to be) the highly directional 
f-electrons in actinides (U here).  The phonon anomaly is observed at 
$q=\frac{1}{2}[\xi,0,0]$.  The martensitic phase is presumably 
monoclinic but this is not yet known although there is a small peak in 
specific heat recently observed \cite{lashley} at 1.2 K.

We notice that for materials in this group, the lock-in transition 
at T$_{II}$ occurs without a change in symmetry (at least on 
an average). 

{\bf (II)} In this second category we include those systems, although they 
exhibit the two modulated phases, in which the lock-in transition is 
accompanied by a change in symmetry.  As a prototype we take the NiTi(Fe) 
material \cite{shapiro84,salamon85,folkins89}.  This shape memory alloy 
(SMA) has a high temperature cubic structure and exhibits an IC phase.  
It also locks into a CM phase with trigonal symmetry (i.e., an intermediate 
R phase) before going into a monoclinic martensite \cite{barsch}. Very 
recently, it has been pointed out that the R phase is a legitimate 
martensite competing with the monoclinic martensite \cite{shindo}.  
Note that this competing martensitic phases scenario is consistent 
with our framework.  There is a small tendency for $C'$ to have an up-turn 
with decreasing temperature\cite{ren99,ren01} and the phonon anomaly is 
observed at $q=\frac{1}{3} [\xi,\xi,0]$.  In this case the freezing of 
the phonon with decreasing temperature (unlike the case I above) is 
accompanied even by a change in space group symmetry (from cubic to 
trigonal) related to the underlying phonon displacement amplitude 
\cite{shindo}.  Note that stoichiometric AuCd (Refs. 24,28) also has 
a B2 to R phase transition similar to NiTi(Fe).   

{\bf (III)} In this category we include systems in which the CM phase is 
suppressed (T$_{II}$=T$_M$). This is the case of the  Ni$_{x}$Al$_{1-x}$
($.45<x<.63$) (Refs. 29,30), Fe$_{1-x}$Pd$_x$ ($x<.32$) (Refs. 31,32), 
Fe$_3$Pt(ordered) (Ref. 33) and In-Tl (Ref. 34) materials.  They exhibit 
an IC phase (referred to as tweed, in real space) but the CM phase is 
absent before they undergo a MT.  The phonon anomaly in NiAl is 
observed at $q\simeq\frac{1}{6}[\xi,\xi,0]$ with a monoclinic 
martensitic phase.  

{\bf (IV)} In the case of Cu-based shape memory alloys (e.g. Cu-Zn-Al), 
no modulated phase is observed. There is a unique  phase transition from 
the high temperature cubic austenite phase directly to the martensitic 
phase.  In phonon dispersion curves there is almost no dip but 
significantly, the whole (TA$_2$) phonon branch has a quite low energy.  
These materials can be termed as ordinary SMA with entropy driven   
\cite{planes01,msf} MT.   

We note that materials discussed in the above categories exhibit the 
same premonitory effects but different structural precursors 
\cite{precursors}.  The different transitions are accompanied by 
anomalies in various physical quantities with different magnitudes. 

{\bf (i)} Softening of a specific phonon branch accompanied by a dip.  
Figure 2 shows experimental examples for Ni-Mn-Ga (Refs. 12,13) 
$\alpha$-U (Ref. 36) and Ni-Al (Ref. 30). For the purpose of 
discussion below, the two cubic alloys Ni-Mn-Ga and Ni-Al are shown 
together and $\alpha$-U (orthorhombic) is shown in the inset.  Note the 
appealing similar behavior of phonons in Ni-Mn-Ga and $\alpha$-U. 
Except for a linear decrease, there is no interesting feature in Ni-Al 
phonons such as a conspicuous change in slope. The different transitions 
are denoted by arrows. The increase in energy at low temperature for 
Ni-Mn-Ga is an indication for the phonon freezing at T$_{II}$. 
Presumably the $\alpha$-U should also exhibit similar low temperature 
increase in phonon energy but unfortunately the experimental data is lacking.

{\bf (ii)} Softening of the relevant elastic constants (i.e., the 
long-wavelength 
limit of TA$_2$ phonon branch).   An example for $C'$ and $C_{44}$ 
softening as the temperature is decreased to T$_{II}$ and elastic 
constant  `hardening' upon subsequent cooling\cite{manosa97,stenger98} 
to T$_M$ for Ni-Mn-Ga is depicted in Fig. 3(a).  In contrast, other 
than a linear decrease there is no structure in the behavior of elastic 
constants for Ni-Al, also shown on the same figure for comparison.  In 
Fig. 3(b) we show the temperature variation of three of the relevant  
orthorhombic elastic constants for $\alpha$-U (Ref. 37).  There is a 
striking similarity in the $C_{11}$ and $C_{44}$ variation for 
$\alpha$-U and the $C'$ and $C_{44}$ variation for Ni-Mn-Ga (Ref. 38).  
Unfortunately, the $C_{12}$ low temperature data for $\alpha$-U is 
lacking.  We also note that a somewhat similar incipient anomaly is 
observed in NiTi(Fe) (Ref. 39).  In this work, we shall argue that 
the up-turn (or elastic constant `hardening') with decreasing 
temperature at T$_{II}$ is characteristic of systems undergoing 
multi-stage modulated structural transformations. This point will be 
discussed below.

{\bf (iii)} Anomalies in the specific heat, resistivity, magnetic 
susceptibility and other thermodynamic variables are also observed. 
Examples include  Ni-Mn-Ga (Ref. 8), NiTi(Fe) (Ref. 22), $\alpha$-U 
(Ref. 15) and other actinides \cite{hecker}.

\begin{center}
\begin{figure}[th]
\includegraphics[width=8.6cm,clip]{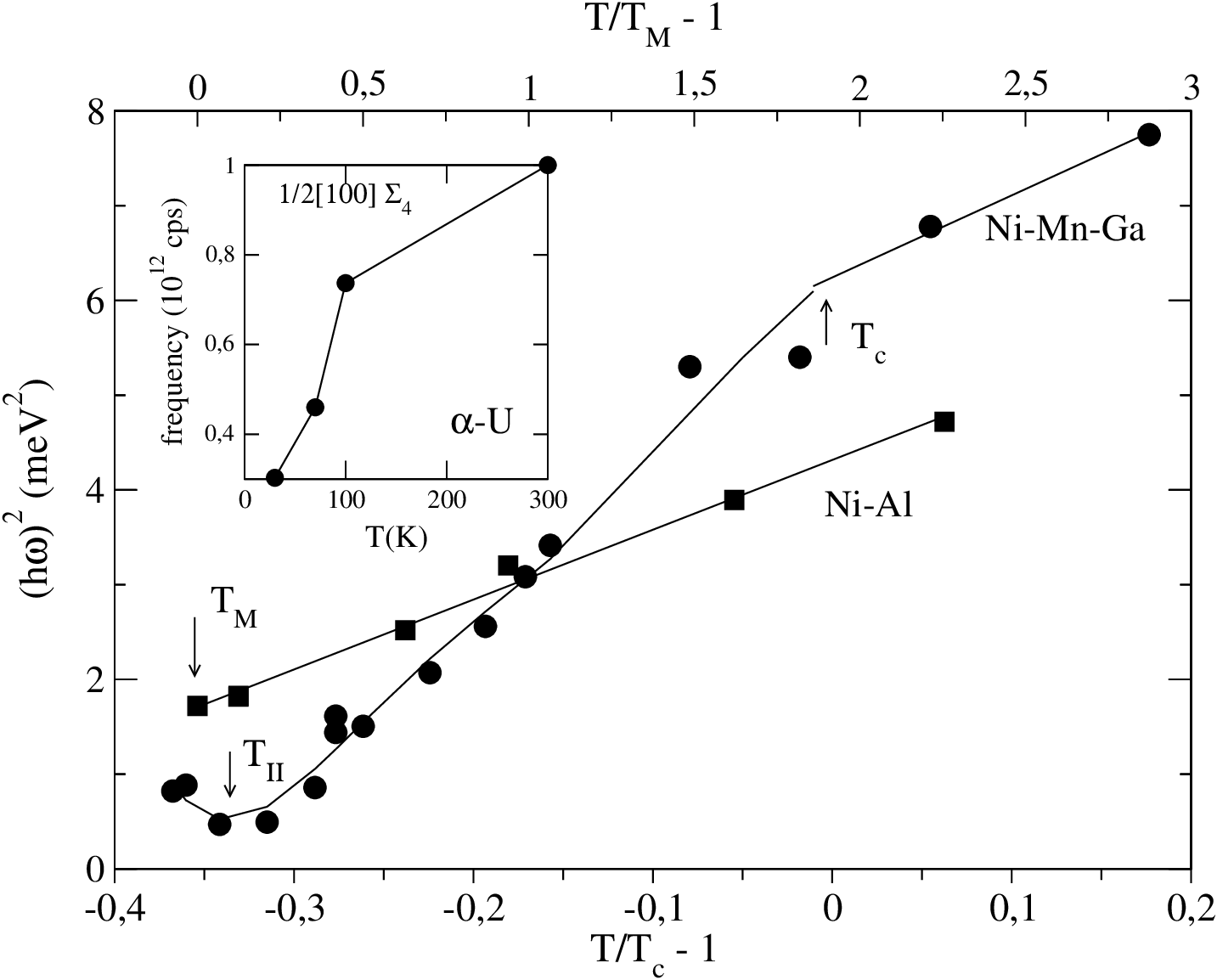}
\caption{Energy of the anomalous phonon as a 
function of reduced temperature for Ni-Al (T$_M$) and Ni-Mn-Ga (T$_c$).
T$_c$ is the Curie temperature and T$_I$ cannot be properly identified.  
The inset shows the phonon softening in $\alpha$-U. Data have been extracted
from references indicated in the text.}
\end{figure}
\end{center}

The IC phase is detected by diffuse satellite reflections that appear 
at incommensurate positions \cite{tanner}.  Examples include Ni-Mn-Ga 
(Ref. 10), Ni-Al (Ref. 29), Fe-Pd (Ref. 32), NiTi(Fe) 
(Ref. 42). It is not caused by a phonon instability but it is due 
to local inhomogeneities \cite{kartha95} (e.g., compositional 
fluctuations, crystal defects, residual strain) that (locally) couple 
to the soft modes. The emerging phase is thermodynamically stabilized 
by anisotropic, long-range elastic forces \cite{shenoy}.  However, 
this remains an open question for future investigation.  The further 
lock-in (or freezing) of the 
phonon at commensurate positions requires an additional degree of 
softening of the anomalous phonon frequency with decreasing 
temperature before it reaches the martensitic transition T$_M$.   
It is clear from Fig. 2 that in the case of Ni-Mn-Ga, the 
magnetism provides (through a magnetoelastic coupling \cite{TCV}) the 
enhancement of softening necessary for freezing.  
\begin{center}
\begin{figure}[th]
\includegraphics[width=6.0cm,clip]{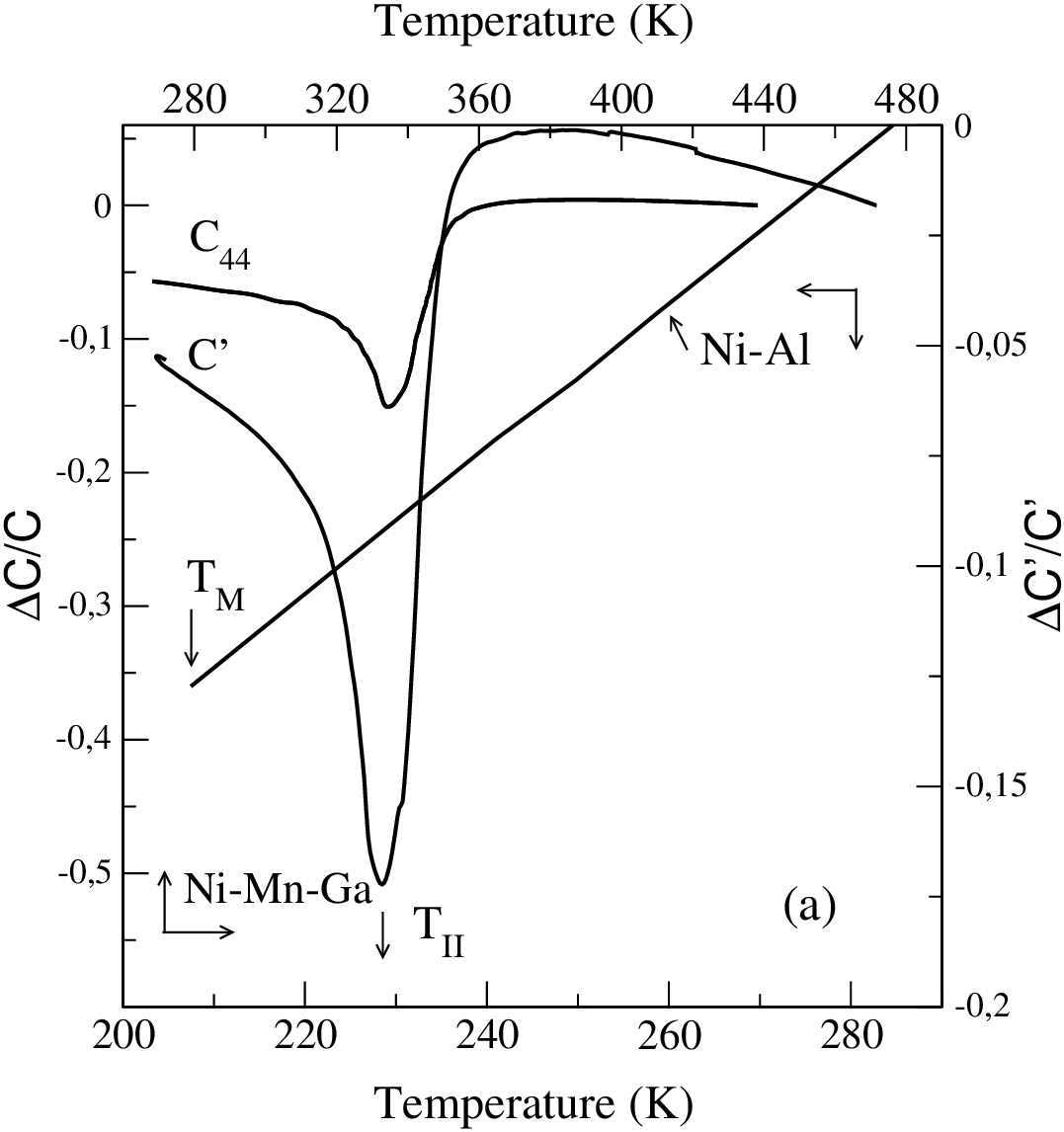}
\includegraphics[width=6.0cm,clip]{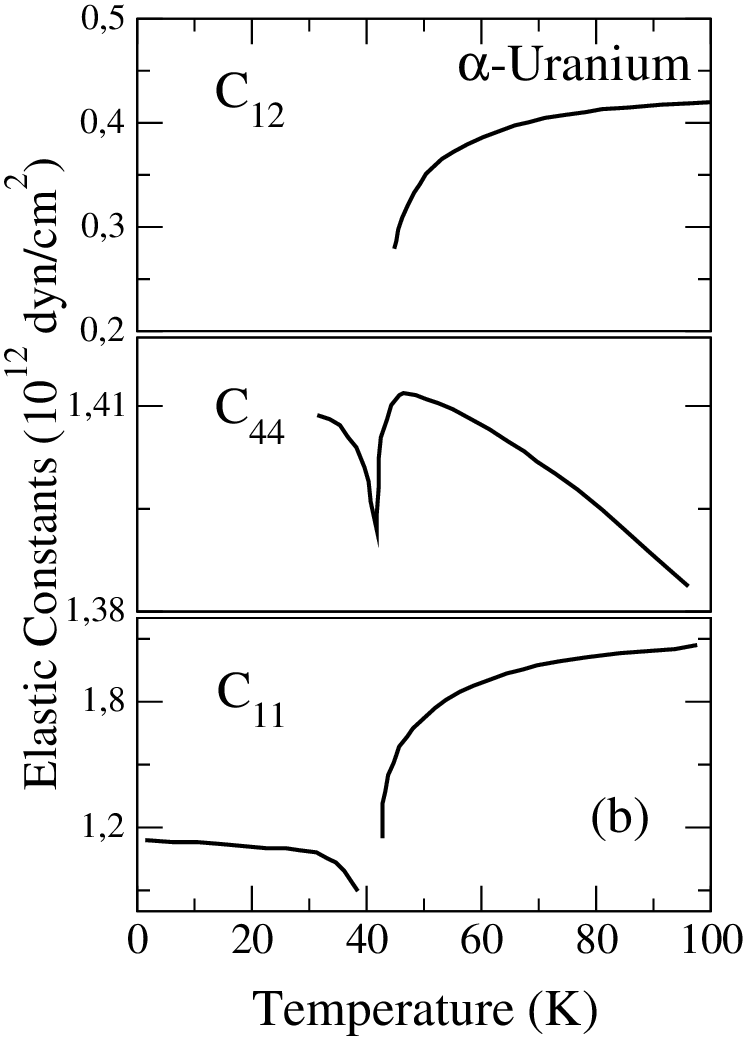}
\caption{(a) Elastic constant anomaly in Ni-Mn-Ga compared to that in
Ni-Al. (b) Elastic constant anomaly in $\alpha$-Uranium. References from
where data have been extracted are indicated in the text.}
\end{figure}
\end{center}
Moreover, it is known that in many martensitic alloys the transition 
temperature T$_M$ is very sensitive to the electron concentration per 
atom (e/a).  This has never been quantified, but clearly so.  In Fig. 
4(a) we compare various transition temperatures for three materials 
\cite{shapiro84}: NiTi and NiTi(Fe) for two different compositions as 
a function of e/a. [The stable phase for $T_{II}<T<T_M$ is rhombohedral 
(R) and modulated \cite{shindo}].  This dramatic dependence of lattice 
stability with the e/a ratio is also observed in Ni-Mn-Ga (Ref. 8) and 
Fe-Pd (Ref. 44) as well [see Fig. 4(b)].  However, the trend is reversed 
in Fe-Pd [inset of Fig. 4(b)] because, unlike Ni-Mn-Ga and Ni-Ti, Fe-Pd 
is a close-packed (magnetic) structure.  Small changes in the relative 
alloying percentages of the elements may produce significant variations 
in T$_M$ and therefore in the observed behavior of the materials. This 
is especially relevant in Cu-based alloys and actinides, particularly 
in Ga-stabilized \cite{hecker} Pu.  Beyond a certain alloying 
percentage, the martensitic transformation can be arrested totally 
(e.g., above \cite{muto88,muto90} 32 \% Pd in Fe-Pd).  More precisely, 
as T$_M$ increases, the freezing of the phonon prior to the martensitic 
transition becomes less likely.
\begin{center}
\begin{figure}[th]
\includegraphics[width=6.0cm,clip]{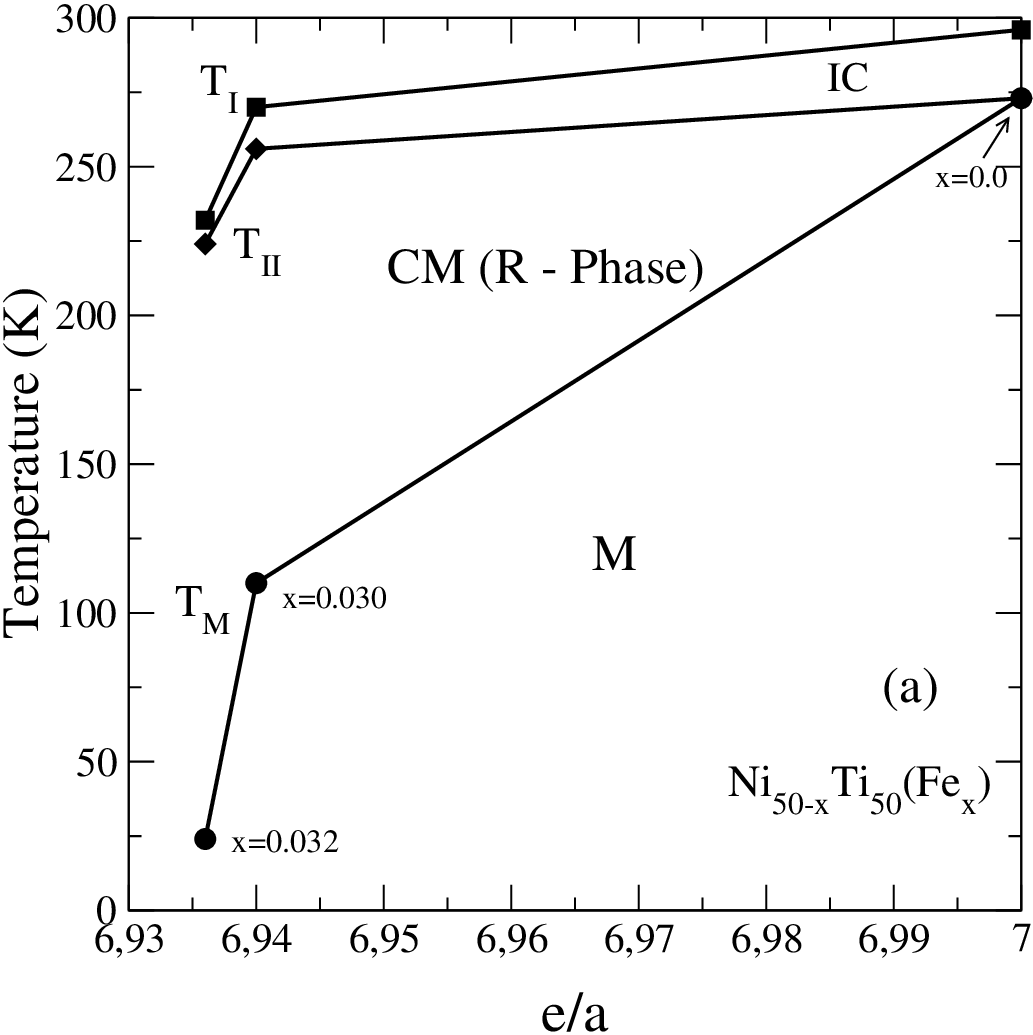}
\includegraphics[width=6.0cm,clip]{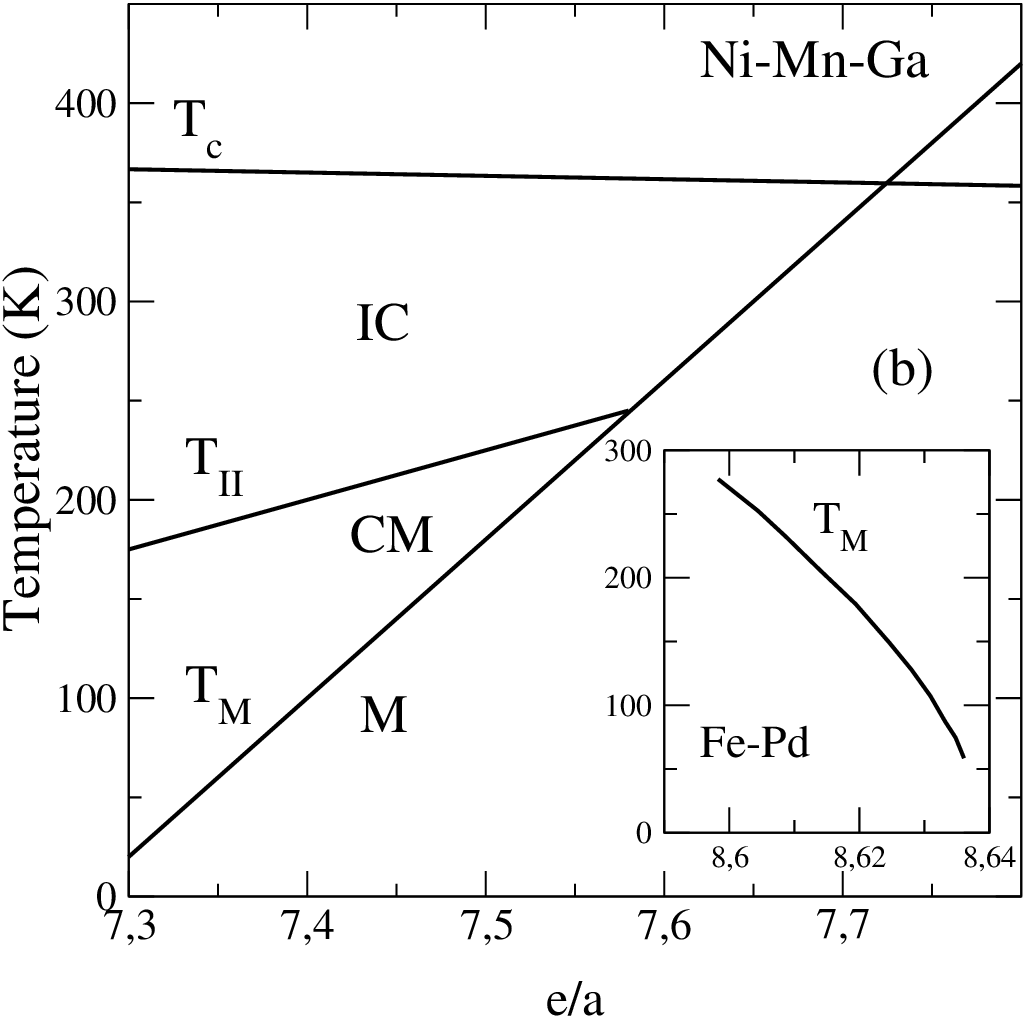}
\caption{Various transition temperatures as a function of e/a in 
(a) Ni-Ti (after Ref. 21), (b) Ni-Mn-Ga (Ref. 8) and in the inset Fe-Pd 
(Ref. 44).}
\end{figure}
\end{center}

\section{Effective 1D model} 
From the full three-dimensional crystal symmetry analysis we have a 
Ginzburg-Landau model in terms of all (six) strain tensor components 
\cite{vasilev,rasmussen}, all shuffle \cite{folkins89} order parameters  
and the magnetization \cite{vasilev} with the symmetry allowed couplings 
between strain and shuffle, strain and magnetization as well as shuffle 
and magnetization.  For specific strain components and the shuffle in a 
particular direction we can obtain an effective one-dimensional analog 
of the full 3D model.  We emphasize that our goal here is to address 
the ion-displacive (or phonon-driven \cite{ion}) anomalies (above T$_M$) 
but not the microstructural aspects. 

We focus on the $\vec{q}=[\xi \xi 0]$ projection of the
cubic$\rightarrow$tetragonal distortion. The relevant order parameters 
are the amplitude of the
anomalous phonon $\eta$ and the tetragonal strain $e$ ($e_3$ in standard
symmetry-adapted notation). The simplest Landau free-energy expansion is:
\begin{equation}
F(e,\eta) = \frac{{\omega}^2}{2} \eta^2 + \frac{\beta}{4} \eta^4 +
\frac{\gamma}{6} \eta^6 + \frac{C'}{2} e^2 + \frac{A}{3} e^3 + \frac{B}{4} e^4
+ \kappa_{e \eta} e \eta^2,
\label{energy1}
\end{equation}  
where $\omega$ is the frequency of the anomalous phonon and $C'$ the elastic
constant defined as $C' = (C_{11} - C_{12})/2$. Both soften with decreasing
temperature (for the cubic materials considered here). Here $\kappa_{e\eta}$ 
is a material dependent parameter denoting symmetry allowed coupling 
between strain and shuffle. 

For the analysis of the modulated phases, we write an effective free-energy in
terms of the shuffle $\eta$ only; i.e.,
\begin{equation}
F_e(\eta)= \frac{{\omega}^2}{2} \eta^2 +
\frac{\beta_r}{4} \eta^4+ \frac{\gamma}{6} \eta^6,
\end{equation}
with the renormalized coefficient (including fluctuations and anharmonicities 
\cite{jak92}) being
\begin{equation}
\beta_r = \beta - {2(\kappa_{e\eta})^2}/{C'}.
\end{equation}
The associated
tetragonal distortion is given by \cite{PAL} 
\begin{equation}
e= - (\kappa_{e\eta}/C') \eta^2,
\end{equation}
considering only the harmonic terms in strain.

In the IC phase $\beta_r>0$ and the last term in the  $\eta$-expansion (2)
is not relevant (in addition to the shuffle amplitude being very small). 
One  possible origin of this phase is local inhomogeneities that couple 
to the soft modes \cite{kartha95} and renormalize the quadratic coefficient 
so that in some regions $\omega^2_r$ becomes negative. $T_I$ is the 
temperature at which long-range elastic forces stabilize the IC phase. 
Then, the intensity of the satellite reflections continuously increases 
from zero with decreasing temperature and consequently the magnitude of 
the associated tetragonal strain, which serves as a precursor to martensite.

As the order parameter $\eta$ increases \cite{manosa01,salamon85}, the
sixth-order term in (2)
becomes important. The lock-in of the phonon at commensurate 
positions requires that $(\kappa_{e\eta})^2/C'$ be large enough so that 
 $\beta_r<0$.  This is 
a phonon instability and the transition at $T_{II}$ is of first-order.

Subsequent cooling is dominated by the increasing phonon amplitude and 
the softening of $C'$ is not required for the structural instability at 
T$_M$. Actually, materials undergoing the lock-in transition (Ni-Mn-Ga, 
$\alpha$-U) exhibit a {\it hardening} of the relevant elastic constant(s) 
in the range $T_M<T<T_{II}$ (see Fig. 3).  We suggest this unusual behavior 
(when approaching a structural phase transition) is {\it characteristic} 
of materials for which the phonon instability occurs {\it prior} to the 
structural instability. The {\it up-turn} at $T_{II}$ reflects the higher 
stability of the CM (averaged cubic) phase with respect to the parent 
cubic phase. The further structural instability results from a symmetry 
allowed coupling of strain to the increasing value of $\eta$ [that is, 
from Eq. (4), $e= - (\kappa_{e\eta}/C') \eta^2$].

This analysis may be expanded to include secondary couplings. 

(i) Effect of softening in $C_{44}$: Ren and Otsuka \cite{RenOtsuka}
pointed out that in some materials $C_{44}$ also softens with temperature 
and consequently there may be a competition between both $C'$ and $C_{44}$ 
modes. In that case:
\begin{equation}
F(e,\epsilon,\eta) = F(e,\eta)+ \frac{C_{44}}{2} \epsilon^2 +
\kappa_{\epsilon\eta} \epsilon \eta^2,
\end{equation} 
where $\epsilon$ is the corresponding (symmetry-adapted shear $e_4$) strain and
$\kappa_{\epsilon\eta}$ the strength of the symmetry allowed coupling term. 
The expression of F(e,$\eta$) is given by (\ref{energy1}). We may write
an effective free-energy in terms of $\eta$ and obtain that the renormalized
coefficient is now given by: 
\begin{equation}
\beta_r = \beta - \frac{2(\kappa_{e\eta})^2}{C'}\left[1 +
 \frac{ (\kappa_{\epsilon\eta}/\kappa_{e\eta})^2}{A} \right],
\end{equation}
where $A=C_{44}/C'$ is the elastic anisotropy. The coupling between both shear
modes is then given by:
\begin{equation}
\epsilon = \left[\frac{\kappa_{\epsilon\eta}}{\kappa_{e\eta}} \right
]\frac{e}{A}. 
\end{equation}
We emphasize that this relationship between the two shear strains is 
naturally mediated through the elastic constant anisotropy ($A$). 

(ii) Role of magnetism:  From the existing data, it appears to renormalize 
the shuffle coefficients (see Fig. 2).  Clearly, d-electrons are responsible 
in NiTi(Fe) and Ni-Mn-Ga while f-electrons are crucial in $\alpha$-U and 
other actinides, especially \cite{hecker,kotliar} Pu.  The `kink' observed 
at the Curie point $T_c$ is described by writing  $\omega_{r}^{2}(m) = 
\omega^2 + \kappa_{m\eta} m^2$ (with $\kappa_{m\eta}<0$), where $m$ denotes 
magnetization. Other couplings that renormalize the strain coefficients are 
also very likely. Unfortunately there is no data on the 
behavior of $C'$ around $T_c$.

(iii) For $\alpha$-U it follows from symmetry analysis that the monoclinic 
strain is coupled to the frozen phonon amplitude \cite{walker}. Thus, the 
anomaly at 1.2 K in the specific heat \cite{lashley} may be an indication 
of MT to a monoclinic phase.  This may also happen for other actinides, 
including the apparently strange behavior \cite{hecker,kotliar} of Pu.

\section{Conclusion} 
We have proposed a general framework for understanding {\it multi-stage} 
martensitic transformations which is consistent with a large 
amount of experimental data on different materials such as Ni-Al, 
Ni-Mn-Ga, NiTi(Fe), $\alpha$-U, Fe-Pd, Au-Cd, etc.  Specifically, 
presence of an incommensurate, and in some cases an additional 
commensurate, modulated phase as precursors to the MT reflects itself in the anomalous (softening and 
further `hardening') of both the elastic constants and phonons.  In the 
case of some magnetic martensites, particularly Ni-Mn-Ga, the 
phonon softening is enhanced by magnetism.   We propose additional 
experiments that may test the broader validity of our scenario.  
(i) Measurements of $C'$ in Ni-Mn-Ga as a function of temperature 
with the hope of observing a change in slope around $T_c$. (ii) 
Low temperature (below 30 K) measurements of phonon dispersion in 
$\alpha$-U to observe a possible up-turn (akin to that in Ni-Mn-Ga). 
(iii) Measurements indicating a possible structural phase transition 
to a monoclinic martensite around 1.2 K in $\alpha$-U. 
 
From the new perspective presented here it is clear that, concerning 
structural behavior, the underlying physics in such disparate 
systems as shape memory alloys and $\alpha$-U is quite similar.  
We have attempted to extract some unifying principles that provide 
new insight into multi-stage, modulated structural transformations. 
Moreover, these connections may enable exchange of ideas and expertise 
from one set of materials to another and vice versa. 

Inclusion of gradient terms will allow us to study domain walls, 
especially antiphase boundaries, twin boundaries and microstructure 
by augmenting the Landau free energy with Ginzburg terms \cite{rasmussen} 
in multi-stage transformations.  It remains to be explored how the 
multi-stage character of the transformation affects domain wall 
orientation and energetics, specifically the microstructure.   

This global view of the many factors implicated for different materials 
has drawn our attention particularly to the need for further study, in 
any complex material, for new, refined measurements of heat capacities 
and magnetic susceptibility. 
                                                      
\section{Acknowledgment} 
We thank L. Ma\~nosa, E. Vives and J.C. Lashley for fruitful discussions. 
We are indebted to G.R. Barsch, J.A. Krumhansl, K. Otsuka and S.M. Shapiro 
for critical comments.  A.S.  gratefully acknowledges a fellowship from 
Iberdrola (Spain).  This work was supported in part by the U.S. Department 
of Energy and in part by the CICyT (Spain) project MAT2001-3251 and CIRIT 
(Catalonia) project (2001SGR00066).


\begin{references}
\bibitem{precursors} By precursors we specifically mean structural effects
  (e.g. modulated phases) above T$_M$ while anomalies in phonons, elastic
  constants and others, above T$_M$, are refered to as premonitory effects.
\bibitem{dataU} For careful measurements of the phonon energy below 30 K 
we expect an up-turn in $\alpha$-U similar to that for Ni-Mn-Ga.
\end{references}
\end{document}